%% file: paper.v1.tex
\documentclass[twocolumn]{aastex63}

\usepackage{amssymb}
\usepackage{amsmath}
\usepackage{amsfonts}
\usepackage{graphicx}
\usepackage{color}
\usepackage{hyperref}

\input{defs.tex}

\allowdisplaybreaks

\usepackage[normalem]{ulem}

\begin{document}

\title{Acoustic wake in an isothermal profile: dynamical friction and gravitational wave emission}

\author{Gali Eytan}
\email{gali.eytan@campus.technion.ac.il}
\affiliation{Physics Department, Technion -- Israel Institute of Technology, Haifa 3200003, Israel}

\author{Vincent Desjacques}
\email{dvince@physics.technion.ac.il}
\affiliation{Physics Department, Technion -- Israel Institute of Technology, Haifa 3200003, Israel}

\author{Robin Buehler}
\email{robinbuehler@campus.technion.ac.il}
\affiliation{Physics Department, Technion -- Israel Institute of Technology, Haifa 3200003, Israel}

\date{\today}

\begin{abstract}
    We consider the motion of a circularly-moving perturber in a self-gravitating, collisional system with spherically symmetric density profile. We concentrate on the singular isothermal sphere which, despite its pathological features, admits a simple polarization function in linear response theory. This allows us to solve for the acoustic wake trailing the perturber and the resulting dynamical friction, in the limit where the self-gravity of the response can be ignored. In steady-state and for subsonic velocities $v_p<c_s$, the dynamical friction torque $F_\varphi\propto v_p^3$ is suppressed for perturbers orbiting in an isothermal sphere relative to the infinite, homogeneous medium expectation $F_\varphi\propto v_p$. For highly supersonic motions, both expectations agree and are consistent with a local approximation to the gravitational torque. At fixed resolution (a given Coulomb logarithm), the response of the system is maximal for Mach numbers near the constant circular velocity of the singular isothermal profile. This resonance maximizes the gravitational wave (GW) emission produced by the trailing acoustic wake. For an inspiral around a massive black hole of mass $10^6 M_\odot$ located at the center of a (truncated) isothermal sphere, this GW signal could be comparable to the vacuum GW emission of the black hole binary at sub-nanohertz frequencies when the small black hole enters the Bondi sphere of the massive one. The exact magnitude of this effect depends on departures from hydrostatic equilibrium and on the viscosity present in any realistic astrophysical fluid, which are not included in our simplified description.
\end{abstract}

\section{Introduction}

Astrophysical bodies from the size of black holes to galaxies and dark matter halos merge by transferring orbital energy and angular momentum to their gravitational interaction with the astrophysical environment. This physical process is known as Dynamical Friction (DF) since \cite{chandrasekhar:1943}, and is a key ingredient in cosmic structure formation \citep{tremaine/etal:1975,paczynski:1976,murai/fujimoto:1980,binney/tremaine:1987,kauffmann/etal:1993,ibata/lewis:1998,somerville/primack:1999,vandenbosch/etal:1999,ostriker:1999,cole/etal:2000,goldreich/etal:2004,croton/etal:2006,boylankolchin/etal:2008,mo/etal:2010}. In a seminal work, \cite{chandrasekhar:1943} derived an expression for the gravitational drag or DF force produced by a point-like perturber moving in linear motion in a infinite, homogeneous and collisionless medium. His calculation has been extended to different boundary conditions and backgrounds to gain understanding on the physics of DF.
The latter includes stellar distributions \citep{lyndenbell/kalnajs:1972,tremaine/weinberg:1984,palmer/papaloizou:1985,weinberg:1986,bekenstein/maoz:1992,nelson/tremaine:1999,banik/vandenbosch:2021,ginat/etal:2023,kipper/etal:2023}, gaseous media \citep{dokuchaev:1964,ruderman/spiegel:1971,rephaeli/salpeter:1980,just/kegel:1990,ostriker:1999,salcedo/brandenburg:2001,kim/kim:2007,lee/stahler:2011,vicente/etal:2019,salcedo:2019,desjacques/etal:2022} and, more recently, exotic dark matter backgrounds \citep{hui/etal:2017,baror/etal:2019,berezhiani/etal:2019,annulli/etal:2020,chavanis:2020,chavanis:2021,traykova/clough/etal:2021,hartman/etal:2021,Bar:2021jff,rodrigo/cardoso:2022,buehler/desjacques:2023,foote/etal:2023,tomaselli/etal:2023,traykova/vicente/etal:2023,boudon/etal:2024}. In all cases, the DF force is given by the gravitational deceleration produced by a nonlocal or global density wake trailing the perturber, as was already pointed out by \cite{kalnajs:1971,kalnajs:1972}.

In this work, we calculate the global density wake produced by a circularly-moving perturber in an ideal fluid with a $r^{-2}$ isothermal profile. The case of a general, realistic fluid with finite mean free path $\lf$ explored in, e.g., \cite{katz/etal:2019}, is beyond the scope of this paper. Nonetheless, our study complements \cite{tremaine/weinberg:1984,weinberg:1986,kaur/stone:2022}, who focused on collisionless spherical systems, and provides useful insights on dynamical friction for self-gravitating spherical systems with vanishing $\lf$. We assume that the self-gravity of the density wake can be neglected and use linear response theory to compute the polarization function \citep{kalnajs:1971,tremaine/weinberg:1984,binney/tremaine:1987,weinberg:1989,nelson/tremaine:1999}. Following \cite{tremaine/weinberg:1984,desjacques/etal:2022} for the collisionless and collisional limit, respectively, we expand the perturbations in multipoles to solve for the trailing acoustic wake and the friction coefficient. We concentrate on the singular isothermal profile, which admits a simple polarization function at the expense of its well-known pathological features \citep{chandrasekhar:1939}. Our approach can, in principle, be applied to any spherically symmetric profile in hydrostatic equilibrium.

The paper is organized as follows. In \S\ref{sec:theory}, we summarize the assumptions behind our approach and spell out the derivation of the polarization function, the acoustic wake density and the dynamical friction. We validate our approach with the infinite, homogeneous case for which semi-numerical simulations are available. In \S\ref{sec:SIS}, we apply our method to a singular isothermal sphere, and derive the trailing acoustic wake and the dynamical friction in the steady-state regime. In \S\ref{sec:GW}, we discuss the gravitational wave emission emanating from a trailing acoustic wake induced by a compact binary embedded in a (truncated) isothermal sphere. We conclude in \S\ref{sec:conclusion}.

\section{Theory}

\label{sec:theory}

Consider a spherically symmetric, inviscid gaseous medium in hydrostatic equilibrium of density $\rho_0(r)$ and (adiabatic) sound speed $c_0(r)$. The equation of state for the pressure $P_0(r)$ is the polytrope $P_0=K \rho_0^\gamma$ where $K$ is a constant and $\gamma$ is the adiabatic index. The potential of this self-gravitating medium is $\phi_0(r)$. 

We are interested in the li near response of this gaseous system to an external, pointlike perturbation of mass $M$ moving on a fixed  circular orbit of radius $a$ and angular velocity $\Omega>0$, in the $x-y$ plane. 

Let $\vr_p=\vr_p(t)$ and ${\vr_p}'=\vr_p(t')$ be the position of the perturber at "present-day" time $t$ and "retarded" time $t'$, respectively, relative to the center of the density profile.
We have $\vr_p(t) = a(\cos\Omega t\,\xvh+\sin\Omega t\,\yvh)$. The Mach number is $\mach\equiv v_p/c_s$, where $v_p = \Omega a$ is the perturber's circular velocity.
The point mass has a density $\rho_p(\vr,t) = M h(t)\delta^D(\vr-\vr_p(t))$, where the time-dependent function $h(t)$ is unity (zero) when the perturber is turned on (off).

\subsection{Methodology}

\subsubsection{Sound wave equation}

Adding the perturber to the system leads to linearized continuity and momentum conservation equations involving perturbations in the gas density $\delta\rho$ and pressure $\delta P = c_0^2 \delta\rho$ as well as the perturber's gravitational potential $\phi_p$,
\begin{align}
\partial_t\delta\rho &=- \grad\cdot\big(\rho_0 \delta\vv\big) \\
\partial_t\big(\rho_0\delta\vv\big) &= -\grad\delta P - \rho_0 \grad\delta\phi - \delta\rho \grad\phi_0 - \rho_0\grad\phi_p\;.
\end{align}
These equations are supplemented by the perturbed Poisson equations
\begin{equation}
    \nabla^2\delta\phi = 4\pi G \delta\rho \quad \mbox{and}\quad \nabla^2\delta\phi_P = 4\pi G \delta\rho_P \;,
\end{equation}
where $\delta\phi$ is the perturbation to the potential of the gaseous sphere. Using the hydrostatic equilibrium condition $(1/\rho_0)\grad P_0+\grad\phi_0=0$, they can be combined to yield a perturbation equation of the form
\begin{multline}
    \label{eq:soundw1}
    \ddot\delta\rho - \nabla^2\delta P - \grad\phi_0\cdot\grad\delta\rho - 4\pi G\rho_0\delta\rho \\ = \grad\big(\rho_0\grad\phi_p\big) + \grad\big(\rho_0\grad\delta\phi\big)\;.
\end{multline}
The right-hand side involves the total perturbation in the gravitational potential, which is the sum of the perturber's potential $\phi_P$ and the perturbation $\delta\phi$ induced by the trailing density wake. For infinite, homogeneous systems, the right-hand side is simply the total density perturbation $\rho_P+\delta\rho$. For realistic, finite and inhomogeneous systems, there is an additional "non-local" term $\grad\rho_0\cdot\grad(\phi_p+\delta\phi)$ which significantly complicates the analysis. Note also that the contribution of the unperturbed gravitational field $\grad\phi_0$ is neglected in the treatment of the infinite, homogeneous case following the "Jeans swindle". Eq.~(\ref{eq:soundw1}) describes a driven harmonic oscillator, but we shall refer to it as the "sound wave" equation for simplicity.

For the polytropes considered here, the sound wave equation becomes
\begin{equation}
    \label{eq:soundw2}
    \ddot\delta\rho-\nabla^2(c_0^2\delta\rho) - 4\pi G\rho_0\delta\rho-\grad\phi_0\cdot\grad\delta\rho = S(\rho_0,\phi_p,\delta\phi)
\end{equation}
where $S(\rho_0,\phi_p,\delta\phi)=\grad\big(\rho_0\grad(\phi_p+\delta\phi)\big)$ is the source term. We now apply Green's method and look for the Green's function $G(\vr,\vr',t,t')$ which solves
\begin{equation}
    \label{eq:GreenEq}
    \left(\partial_t^2-\nabla^2 c_0^2- 4\pi G\rho_0-\grad\phi_0\cdot\grad\right)G = \delta^D(t-t')\delta^D(\vr-\vr')
\end{equation}
Here, $\grad\phi_0\cdot\grad$ involves only the radial part owing to the spherical symmetry of the unperturbed profile, and it is understood that $\nabla^2$ acts on the product $c_0^2G$. Physically, $G$ is the polarization function \citep[usually denoted by $P$ as in][]{binney/tremaine:1987} since it relates the wake overdensity $\delta\rho$ to the total perturbation.

The polarization function $G$ depends on zeroth order quantities and, therefore, inherits the stationarity (invariance under time-translations) and isotropy (invariance under rotations) of the hydrostatic equilibrium. Therefore, $G$ must be of the form
\begin{equation}
    G(\vr,\vr',t,t') = \sum_{\ell m}\int_\omega g_{\omega\ell}(r,r') e^{-i\omega (t-t')} Y_\ell^m(\rvh) Y_\ell^{m*}(\rvh') \;.
\end{equation}
Here, $g_{\omega\ell}(r,r')$ are the radial components of the polarization function, $Y_\ell^m(\rvh)$ are spherical harmonics,
and the shorthand for the integral over the (circular) frequency of the sound waves $\omega$ is $\int_\omega \equiv \frac{1}{2\pi}\int_{-\infty}^{+\infty}\!d\omega$. Substituting this ansatz into Eq.~(\ref{eq:GreenEq}) and using the representations of the Dirac distributions $\delta^D(t-t')$ and $\delta^D(\vr-\vr')$ in terms of plane waves and spherical harmonics,
\begin{align}
    \delta^D(t-t') &= \int_\omega e^{-i\omega (t-t')} \\
    \delta^D(\vr-\vr') &= \frac{1}{r^2}\delta^D(r-r') \sum_{\ell m} Y_\ell^{m*}(\rvh) Y_\ell^m(\rvh')\;,
\end{align}
the radial Green's equation can be recast into
\begin{multline}
    \label{eq:gwl}
    -r\partial_r^2\Big(r\, c_0^2\, g_{\omega\ell}\Big)- r^2\,\partial_r\phi_0\, \partial_rg_{\omega\ell}- \Big[\Big(\omega^2+c_0^2 k_J^2\Big)r^2 \\ -\ell\big(\ell+1\big)c_0^2\Big]g_{\omega\ell}
    = \delta^D(r-r')
\end{multline}
where $k_J^2(r)=4\pi G\rho_0(r)/c_0^2(r)$ is a (position dependent) Jeans wavenumber.

\subsubsection{Acoustic wake and DF force}

Once the polarization function is known, the (over)density $\delta\rho(\vr,t)$ of the acoustic wake is determined by the (integro-differential) equation
\begin{equation}
    \label{eq:ExactWake}
    \delta\rho(\vr,t) = \int\!\!d^3r' \,\int\!\!dt'\,G(\vr,\vr',t-t')\,
    S(\rho_0',\phi_P',\delta\phi')
\end{equation}
where the primes signify that the source term is evaluated at the coordinates $(\vr',t')$. When the self-gravity of the density wake can be neglected, we have $S\approx \grad(\rho_0\grad\phi_P)$ and the polarization function is equal to the response function of the system:
\begin{equation}
    \label{eq:NoSelfGravWake}
    \delta\rho(\vr,t) \approx \int\!\!d^3r' \,\int\!\!dt'\,G(\vr,\vr',t-t')\,\grad\big(\rho_0'\grad\phi_P'\big)\;.
\end{equation}
For the pointlike perturber considered here, the solution to Eq.~(\ref{eq:NoSelfGravWake}) in terms of the density multipoles $\delta\rho_\ell^m(r,t)$ reads
\begin{multline}
    \label{eq:deltarho}
    \delta\rho_\ell^m(r,t) = 4\pi GM \int_\omega \int\!\!dt'\,e^{-i\omega(t-t')}\, h(t') \\ \times Y_\ell^{m*}(\rvh_p') A_{\omega\ell}(r,r_p') 
\end{multline}
where the function $A_{\omega\ell}(r,r_p')$ is given by
\begin{align}
    \label{eq:Auxiliary}
    A_{\omega\ell}(r,r_p') &= \rho_0(r_p')\,g_{\omega\ell}(r,r_p') \\ 
    &\quad -\frac{1}{2\ell+1}\int\!\!ds s^2\,g_{\omega\ell}(r,s)\partial_s\rho_0(s)\partial_s\!\!\left(\frac{s_<^{\ell}}{s_>^{\ell+1}}\right) \nonumber 
\end{align}
Here, $r_p'=|\vr_p'|$ and $\rvh_p'$ is the unit vector along $\vr_p'$. Furthermore, $s_<={\rm min}(s,r_p')$ and $s_>={\rm max}(s,r_p')$. The acoustic wake pattern contributes a gravitational potential 
\begin{equation}
    \label{eq:PoissonPhi}
    \delta\phi(\vr,t) = -G\int\!\!d^3r' \frac{\delta\rho(\vr',t)}{|\vr-\vr'|}\;,
\end{equation}
the gradient of which gives the DF force experienced by the perturber. Using the multipole expansion of $|\vr-\vr'|^{-1}$, the multipoles $\delta\phi_\ell^m(r,t)$ of this self-gravity perturbation can be recast into the form
\begin{align}
    \label{eq:deltaphi}
    \delta\phi_\ell^m(r,t) &= -\frac{4\pi G}{2\ell+1} \int\!\!ds\,s^2\,\frac{s_<^\ell}{s_>^{\ell+1}}\,\delta\rho_\ell^m(s,t)
\end{align}
where $s_<={\rm min}(s,r)$ and $s_>={\rm max}(s,r)$.

The perturbation $\delta\phi$ generates a gravitational drag force (DF) on the perturber given by
\begin{equation}
    {\bf F}_\text{DF}(t) = F_r(t)\, \rvh(t) + F_\varphi(t)\, \pvh(t) 
\end{equation}
where $\rvh(t)=\vr_p(t)/|\vr_p(t)|$ and $\pvh(t)$ are unit vectors in the radial and tangential direction. By definition, the azimuthal unit vector $\pvh(t)$ is oriented counterclockwise.

The tangential component $F_\varphi(t)$ of the DF force at time $t$ can be obtained from the torque $\vtau(t)$ exerted by the acoustic wake on the perturber:
\begin{align}
    \vtau(t) &= -M \vr_p \times \grad\delta\phi(\vr_p,t) \\
    &=-i M L_z \delta\phi(\vr_p,t)\,\zvh \nonumber  \\
    &\equiv a F_\varphi(t)\,\zvh\;,
\end{align}
where $L_z\equiv -i \partial/\partial\varphi$ is the $z$-component of the angular momentum operator ${\bf L}=-i\,\vr\times\grad$. Therefore, 
\begin{equation}
    \label{eq:Fphi}
    F_\varphi(t) = -i \frac{M}{a} L_z \delta\phi(\vr_p,t)
\end{equation}
which, in essence, is the \cite{lyndenbell/kalnajs:1972} formula for the DF torque.
We shall also consider the radial component given by
\begin{equation}
    \label{eq:Fr}
    F_r(t) = - M \partial_r \delta\phi(\vr_p,t) \;,
\end{equation}
although it is dynamically irrelevant for circular orbits.
These expressions are completely general insofar it applies to any spherically symmetric medium and perturber motion.

\subsection{Consistency check: the homogeneous medium}

The acoustic wake pattern and DF force can be computed easily in the case of steady-state circular motion in an infinite homogeneous background of density $\bar\rho$. This case has been explored in e.g.  \cite{kim/kim:2007,kim/etal:2008,desjacques/etal:2022} \citep[see][for eccentric motion]{buehler/etal:2024,oneill/etal:2024}. Ignoring the self-gravity of the medium (which leads to a frequency shift), the radial Green's equation simplifies to
\begin{equation}
    \label{eq:eqgwlhom}
    -\big(r^2 g_{\omega\ell}'\big)' 
    -\left[\left(\frac{\omega}{c_s}\right)^2r^2-l\big(l+1)\right] g_{\omega\ell} \\
    = \frac{\delta^D(r-r')}{c_s^2}
\end{equation}
which is of the Sturm-Liouville type.
For purely outgoing wave boundary conditions at infinity and a finite response in the limit $r\to 0$, the radial polarization function takes the form
\begin{equation}
    \label{eq:gwlhom}
    g_{\omega\ell}(r,r') = \frac{i}{c_s^2}\bigg(\frac{\omega}{c_s}\bigg)\, j_\ell\!\bigg(\frac{\omega}{c_s}r_<\bigg)\, h_\ell^{(1)}\!\bigg(\frac{\omega}{c_s}r_>\bigg)
\end{equation}
for $\omega\ne 0$.
Here, $r_< = {\rm min}(r,r')$ and $r_>={\rm max}(r,r')$, whereas $j_\ell(x)$ and $h_\ell^{(1)}(x)$ are spherical Bessel and Hankel functions, respectively.
When $\omega=0$, the solution is a powerlaw. We find 
\begin{equation}
    g_{0\ell}(r,r') = \frac{c_s^{-2}}{(2\ell+1)}\frac{r_<^\ell}{r_>^{\ell+1}}
\end{equation}
upon requiring that $g_{0\ell}$ decays to zero at infinity.

\begin{figure*}
    \gridline{\fig{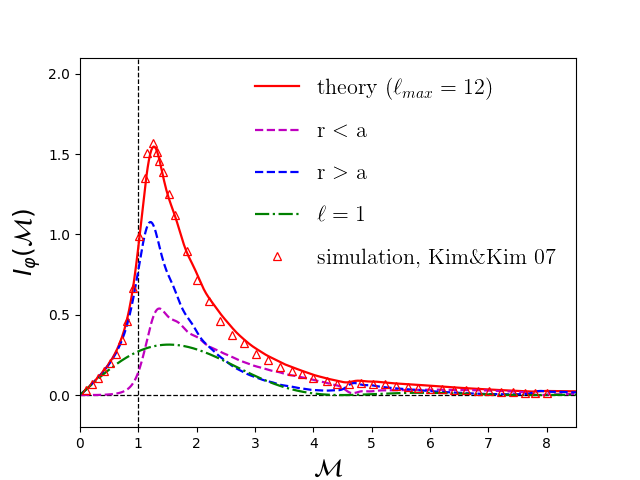}{0.5\textwidth}{}
    \fig{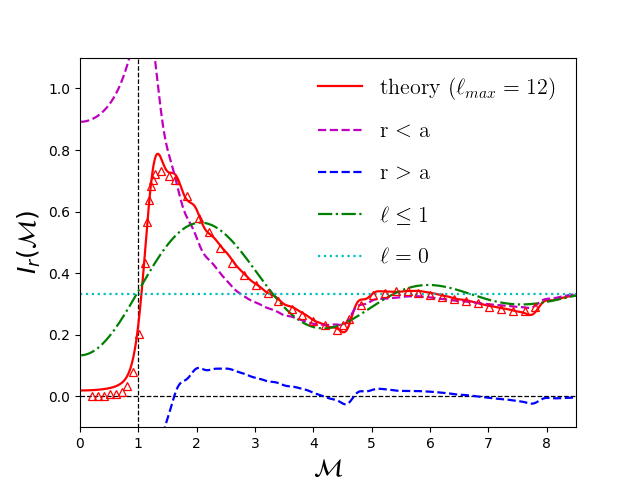}{0.5\textwidth}{}}
    \caption{Tangential (left panel) and radial (right panel) friction coefficient $I_\varphi(\mach)$ and $I_r(\mach)$ as a function of the Mach number $\mach$ in the steady-state regime for a circularly-moving perturber in an infinite homogeneous background. The dashed curves represent the contribution from the inner and outer regions (assuming $\ell_\text{max}=12$), whereas the (cyan) dotted and (green) dot-dashed curves are the contribution from the $\ell=0$ and $\ell\leq 1$ multipoles, respectively. The data is from \cite{kim/kim:2007} (see text for details).}
    \label{fig:hom}
\end{figure*}

On setting $h(t')\equiv~1$ and $\rho(\vr_p')\equiv\bar\rho$ in Eq.~(\ref{eq:deltarho}), the integral over $t'$ simplifies to
\begin{equation*}
    \int\!\!dt'\,e^{-i\omega(t-t')} e^{-im\Omega t'}=2\pi\, \delta^D(\omega-m\Omega)\,e^{-im\Omega t}\;,
\end{equation*}
where the second complex exponential arises from
\begin{equation}
    Y_\ell^{m*}(\rvh_p') = Y_\ell^{m*}\!\left(\frac{\pi}{2},0\right) e^{-im\Omega t'} \;.
\end{equation}
Note that $Y_\ell^0(\pi/2,0)=0$ for $\ell$ even, which reflects the planar symmetry of the response, i.e. $\delta\rho$ is invariant under a reflexion along the $z$ axis.
Inserting this result into the expression of $\delta\rho$ and taking advantage of the fact that $r_p=r_p'=a$ for circular motion, the multipole expansion of the acoustic wake density leads to
\begin{equation}
    \delta\rho_\ell^m(r,t) = 4\pi GM\,\bar\rho\, g_{m\Omega,\ell}(r,a)\, Y_\ell^{m*}(\rvh_p)
\end{equation}
Observe the appearance of all the higher harmonics $m\Omega$ of the fundamental (orbital) frequency. Furthermore, $g_{m\Omega,\ell}(r,a)^* = g_{-m\Omega,\ell}(r,a)$, which ensures that $\delta\rho$ is real. 

The monopole ($\ell=0$) contribution to $\delta\rho$ is independent of the Mach number. It reflects the accretion onto the perturber's orbit and, in steady-state, leads to an overdensity 
\begin{equation}
    \label{eq:drhol0hom}
    \delta\rho_{\ell=0}(\vr,t)\equiv\delta\rho_0^0(r,t)\, Y_0^0(\rvh) =\frac{GM}{c_s^2\,r_>}\bar\rho
\end{equation}
where $r_>={\rm max}(r,a)$. Furthermore, while it does not contribute to the tangential DF force $F_\varphi(t)$, it yields a force in the radial direction which, as we will see shortly, is the dominant contribution to $F_r(t)$ in the limit $\mach\to\infty$.

The multipole coefficients of the perturbation $\delta\phi$ to the medium's gravitational potential follows from Eq.~(\ref{eq:PoissonPhi}) and read
\begin{multline}
    \label{eq:dphihom}
    \delta\phi_\ell^m(r,t) = -\big(4\pi G\big)^2 M \bar\rho\, \big(2\ell+1\big)^{-1} \,Y_\ell^{m*}(\rvh_p) \\ 
    \times \int_0^\infty\!\!ds\,s^2 \frac{s_<^\ell}{s_>^{\ell+1}}\, g_{m\Omega,\ell}(s,a)\;,
\end{multline}
where $s_<={\rm min}(s,r)$ and $s_>={\rm max}(s,r)$. 

Starting from the relation Eq.~(\ref{eq:Fphi}) and using the property $L_z Y_\ell^m = m Y_\ell^m$, and the relations
\begin{align}
    \int_0^1\!\!du\,u^{\ell+2}\, j_\ell(xu) &= \frac{1}{x}\,j_{\ell+1}(x) \\
    \int_1^\infty\!\!du\,u^{-\ell+1}\, h_\ell^{(1)}\!(xu) &= \frac{1}{x}\,h_{\ell-1}^{(1)}\!(x) \nonumber 
\end{align}
which emerge after splitting the integral over $s$ into the intervals $[0,a]$ and $[a,\infty[$, the tangential DF force can be recast into
\begin{equation}
    \label{eq:FDFphisteady}
    F_\varphi(t) = -4\pi\left(\frac{GM}{\Omega a}\right)^2 \bar\rho\, I_\varphi(\mach)
\end{equation}
where 
\begin{align}
    \label{eq:IMphihom}
    I_\varphi(\mach) &= \mach^2 \sum_{\ell=1}^\infty \sum_{m\ne 0} m\, \frac{(\ell-m)!}{(\ell+m)!}\,\big\lvert P_\ell^m(0)\big\lvert^2 \\
    &\quad \times\Big[j_{\ell+1}(m\mach) h_\ell^{(1)}\!(m\mach)+j_\ell(m\mach)h_{\ell-1}^{(1)}\!(m\mach)\Big] \nonumber 
\end{align}
is the tangential friction coefficient. The first (second) term in the square brackets of Eq.~(\ref{eq:IMphihom}) is the contribution from the acoustic wake inside (outside) the orbital radius.
The monopole does not contribute to $I_\varphi(\mach)$ due to Newton's shell (or Birkhoff's) theorem. The dipole ($\ell=1$) term dominates the friction for $\mach\ll 1$. In addition, for $\ell\geq 1$, all the contributions with $m=0$ vanish because $F_\varphi$ cannot be sourced by a density perturbation with azimuthal symmetry. Finally, in the limit $\mach\to 0$, the tangential friction scales as $I_\varphi(\mach)\sim \mach^3/3$, which implies that the tangential drag is proportional to the perturber's velocity for $v_p\ll c_s$.

The friction coefficient $I_\varphi(\mach)$ is displayed in the left panel of Fig.~\ref{fig:hom} as the solid (red) curve. It is compared to the tangential DF extracted from a numerical schemes implementing the Li\'enard-Wiechert potential \citep{kim/kim:2007} as indicated by the triangles. The agreement is excellent as was already demonstrated in \cite{desjacques/etal:2022}. For supersonic motion, $I_\varphi(\mach)$ is affected by a short distance Coulomb divergence, which is regularized by truncating the series expansion Eq.~(\ref{eq:IMphihom}) at a maximum multipole $\ell_\text{max}$. Choosing $\ell_\text{max}=12$ yields a reasonable fit to the simulation data.

Similarly, the radial DF force can be expressed as
\begin{equation}
    \label{eq:FDFradsteady}
    F_r(t) = -4\pi\left(\frac{GM}{\Omega a}\right)^2 \bar\rho\, I_r(\mach)
\end{equation}
where
\begin{align}
    \label{eq:IMradhom}
    I_r(\mach) &= \frac{\mach^2}{3}+i\mach^2 \sum_{\ell=1}^\infty \sum_{m} 
    \frac{(\ell-m)!}{(\ell+m)!}\,\big\lvert P_\ell^m(0)\big\lvert^2 \nonumber \\
    &\quad \times\Big[(\ell+1)\,j_{\ell+1}(m\mach)\, h_\ell^{(1)}\!(m\mach) \nonumber \\ 
    &\qquad -\ell\, j_\ell(m\mach)\, h_{\ell-1}^{(1)}\!(m\mach)\Big]
\end{align}
is the "radial friction coefficient". The term $\mach^2/3$ arises from the monopole Eq.~(\ref{eq:drhol0hom}) of the density perturbation $\delta\rho(\vr,t)$ which gives rise to a radial force
\begin{equation}
    F_r(t)\supset - \frac{GM}{a^2} \left(\frac{4\pi}{3}\frac{GM}{c_s^2 a}\bar\rho a^3\right)=-4\pi\left(\frac{GM}{\Omega a}\right)^2 \bar\rho\,\frac{\mach^2}{3}
\end{equation}
as follows immediately from Newton's shell theorem. Furthermore, when $m=0$, the terms in the square brackets of Eq.~(\ref{eq:IMradhom}) should be replaced by 
\begin{align}
(\ell+1)\,j_{\ell+1}\, h_\ell^{(1)} &\to \left(-\frac{i}{4}\right)\frac{\big(\ell+1\big)}{\left(\ell+\frac{3}{2}\right)\left(\ell+\frac{1}{2}\right)} \\
\ell\,j_\ell\, h_{\ell-1}^{(1)} &\to \left(-\frac{i}{4}\right)\frac{\ell}{\left(\ell+\frac{1}{2}\right)\left(\ell-\frac{1}{2}\right)}\nonumber 
\end{align}
for practical evaluation.

For a static perturber ($\mach=0$), the frequencies are $\omega=m\Omega=0$ so that the density multipoles are given by
\begin{equation}
    \delta\rho_\ell^m(r,t)=4\pi GM\bar\rho\,g_{0\ell}(r,a)\,Y_\ell^{m*}(\rvh_p)\;,
\end{equation}
which yields an overdensity
\begin{equation}
    \delta\rho(\vr) = \frac{GM}{c_s^2}\bar\rho\,\frac{1}{\big|\vr-\vr_p\big|} \qquad (\mach=0)
\end{equation}
that exceeds unity within the Bondi radius $2GM/c_s^2$ of the perturber (where linear response theory becomes invalid).
This density profile originates from the accretion onto the perturber, which is present regardless of the perturber's motion. Since it is isotropic around the constant perturber position $\vr_p$, DF vanishes in the limit $\mach\to 0$. Therefore, the $\mach^2$ term in the multipole expansion of $I_r(\mach)$ must also vanish in the same limit. This translates into a non-trivial constraint among the different multipoles. Furthermore, a series expansion around $\mach=0$ reveals that $I_r(\mach)\propto\mach^4$ for $\mach\ll 1$. The absence of a $\mach^3$ term originates from the fact that, in the limit $v_p\ll c_s$, the perturber can be thought of as moving on a (very short) straight line trajectory on which there is no perpendicular DF force \citep[see also the discussion in][]{desjacques/etal:2022}.

The friction coefficient $I_r(\mach)$ is shown in the right panel of Fig.~\ref{fig:hom}. 
In the limit $\mach\gg 1$, the $m=0$ contribution to $I_r(\mach)$ dominate because the perturber moves so fast that it essentially behaves like a one-dimensional, infinitely thin ring of matter with density $M/2\pi a$, whence the absence of any azimuthal dependence. The dominant contribution originates from the monopole. In the limit $\mach\to 0$, the vanishing of $I_r(\mach)$ is not perfect because multipoles up to $\ell_\text{max}=12$ solely have been included in order to match the resolution of the numerical data.

We have numerically checked that $I_\varphi(\mach)$ and $I_r(\mach)$ precisely match the imaginary and real part, $\Im\big(I(\mach)\big)$ and $\Re\big(I(\mach)\big)$, of the (complex) friction coefficient $I(\mach)$ calculated by \cite{desjacques/etal:2022}. However, our result provides explicit expressions for the contributions from the $r<a$ and $r>a$ region: they are given by the first and the second term in the square brackets of Eqs.~(\ref{eq:IMphihom}) and (\ref{eq:IMradhom}).

\section{The singular isothermal sphere}

\label{sec:SIS}

We now turn to the singular iso\-thermal sphere (SIS) with $\gamma=1$ and $K=c_s^2$, for which the unperturbed density, potential and pressure are
\begin{align}
    \label{eq:SIS}
    \rho_0(r) &= \frac{c_s^2}{2\pi G}\, r^{-2}\\
    \phi_0(r) &= 2 c_s^2\, \ln(r) + \mbox{const} \nonumber 
\end{align}
and $P_0(r) = c_s^2 \rho_0(r)$. This yields a Jeans wavenumber $k_J^2=2/r^2$.
The Green's equation is more complex in this case but, as we shall see below, the methodology is the same as in the infinite, homogeneous case. 

\subsection{Polarization function}

For the SIS profile, the radial Green's equation becomes
\begin{multline}
    \label{eq:eqgwliso}
    -r^2 g_{\omega\ell}''-4 r g_{\omega\ell}'-\bigg[\left(\frac{\omega}{c_s}\right)^2r^2-\ell(\ell+1)+2\bigg]\, g_{\omega\ell} \\
    = \frac{\delta^D(r-r')}{c_s^2}
\end{multline}
The constancy of the sound speed implies $g_{\omega\ell}(\lambda r,\lambda r') = \lambda^{-1} g_{(\lambda\omega)\ell}(r,r')$ for any $\lambda>0$.
The solution to the homogeneous equation (that is, for $r\ne r'$) reads
\begin{equation}
g_{\omega\ell}=\frac{A}{r}\, j_\ell\!\left(\frac{\omega}{c_s}r \right)+\frac{B}{r}\,y_\ell\!\left(\frac{\omega}{c_s}r\right)\;.
\end{equation}
To construct the radial polarization function, we impose the same boundary conditions as in the infinite homogeneous case.
Since $y_\ell$ diverges at the origin, $B$ must be to zero for $r<r'$. Furthermore, the outgoing wave boundary condition at infinity implies that the solution is proportional to $h_\ell^{(1)}$ for $r>r'$. Hence, we have 
\begin{equation}
    g_{\omega\ell}(r,r')=
    \begin{cases}	
    A(r')\, r^{-1}\, j_\ell\!\left(\frac{\omega}{c_s}r\right) & (r<r') \\B(r')\,r^{-1}\,h_\ell^{(1)}\!\left(\frac{\omega}{c_s}r\right) & (r>r')
    \end{cases} \;.
\end{equation}
The continuity and jump conditions at $r=r'$ determine $A(r')$ and $B(r')$ such that, for $\omega\ne 0$, the radial polarization function is given by
\begin{equation}
    \label{eq:gwliso}
    g_{\omega\ell}(r,r')=\frac{i}{c_s^2}\,\left(\frac{\omega}{c_s}\right)\frac{r'}{r}j_\ell\!\bigg(\frac{\omega}{c_s}r_< \bigg)\, h_\ell^{(1)}\!\bigg(\frac{\omega}{c_s}r_>\bigg)
\end{equation}
Except for the asymmetric term $r'/r$ arising from the radial dependence of the unperturbed density profile, it is identical to the polarization function Eq.~(\ref{eq:gwlhom}) of infinite, homogeneous systems. Therefore, we have
\begin{equation}
    \label{eq:gwliso0}
    g_{0\ell}(r,r') = \frac{c_s^{-2}}{(2\ell+1)}\frac{r'}{r}\frac{r_<^\ell}{r_>^{\ell+1}}
\end{equation}
in the special case $\omega=0$.

\begin{figure}
\includegraphics[width=9cm]{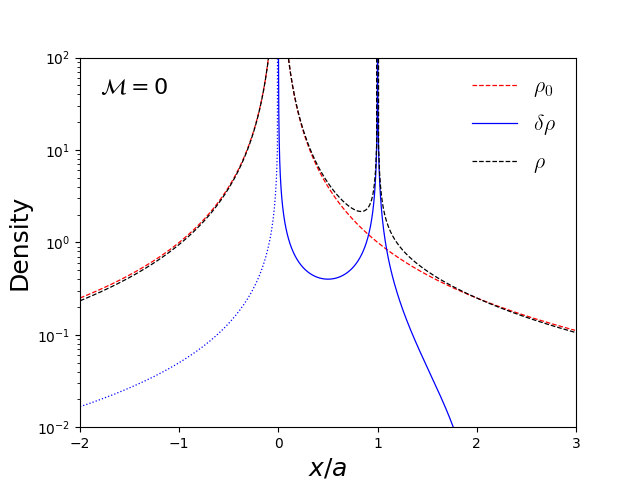} 
\caption{Slice through the density profile of the acoustic wake for a static perturber ($\mach=0$). The unperturbed density is normalized to $\rho_0(a)=1$. A value of $GM/c_s^2 a=0.1$ is assumed. Dotted (blue) lines indicate negative values (of $\delta\rho$). }
\label{fig:WakeZeroMach}
\end{figure}

\subsection{Acoustic wake}

In the steady-state limit, the multipole coefficients of the wake (over)density are given by
\begin{equation}
    \label{eq:drhoSIS}
    \delta\rho_\ell^m(r,t) = 4\pi GM\,  A_{m\Omega,\ell}(r,a)\, Y_\ell^{m*}(\rvh_p) 
\end{equation}  
where $A_{m\Omega,\ell}(r,a)$ is the auxiliary function, Eq.~(\ref{eq:Auxiliary}) with $\omega= m\Omega$ and $r_p'\equiv a$. Introducing 
\begin{align}
    \label{eq;faux}
    \mathcal{J}_\ell^n(x;a,b) &=\int_a^b\!\!du\,u^n\,j_\ell(xu) \\
    \mathcal{H}_\ell^n(x;a,b) &=\int_a^b\!\!du\,u^n\, h_\ell^{(1)}\!(xu) \nonumber
\end{align}
for shorthand convenience, $A_{m\Omega,\ell}(r,a)$ is given by
\begin{align}
    \label{eq;rAwake}
    A_{m\Omega,\ell}(r,a) &=\rho_0(a)\frac{i m\mach}{c_s^2 r}\bigg\{ j_\ell\!\Big(m\mach\frac{r}{a}\Big)\, h_\ell^{(1)}\!(m\mach) \\
    &\quad + \frac{2}{2\ell+1}\bigg[\ell\, h_\ell^{(1)}\!\Big(m\mach\frac{r}{a}\Big)\, \mathcal{J}_\ell^{\ell-1}\!\Big(m\mach;0,\frac{r}{a}\Big) \nonumber \\
    &\quad +\ell\, j_\ell\Big(m\mach\frac{r}{a}\Big)\, \mathcal{H}_\ell^{\ell-1}\!\Big(m\mach;\frac{r}{a},1\Big) \nonumber \\
    &\quad -\big(\ell+1\big)j_l\Big(m\mach\frac{r}{a}\Big)\,\mathcal{H}_\ell^{-\ell-2}\!\big(m\mach;1,\infty\big)\bigg]\bigg\} \nonumber
\end{align}
for $r<a$, and
\begin{align}
    \label{eq:RAwake}
    A_{m\Omega,\ell}(r,a) &= \rho_0(a)\frac{i m\mach}{c_s^2 r} 
    \bigg\{ j_\ell(m\mach)\, h_\ell^{(1)}\!\Big(m\mach\frac{r}{a}\Big)\\
    &\quad + \frac{2}{2\ell+1}\bigg[\ell\, h_\ell^{(1)}\!\Big(m\mach\frac{r}{a}\Big)\, \mathcal{J}_\ell^{\ell-1}\!\big(m\mach;0,1\big) \nonumber \\
    &\quad -\big(\ell+1\big) h_\ell^{(1)}\!\Big(m\mach\frac{r}{a}\Big)\, \mathcal{J}_\ell^{-\ell-2}\!\Big(m\mach;1,\frac{r}{a}\Big) \nonumber\\
    &\quad -\big(\ell+1\big)j_l\Big(m\mach\frac{r}{a}\Big)\,\mathcal{H}_\ell^{-\ell-2}\!\Big(m\mach;\frac{r}{a},\infty\Big)\bigg]\bigg\} \nonumber
\end{align}
for $r>a$. Eqs~.(\ref{eq:drhoSIS}) -- (\ref{eq:RAwake}) are the equivalent of the density response computed in \cite{weinberg:1986,kaur/stone:2022} for self-gravitating collisionless systems. For the boundary conditions assumed here (steady-state), the acoustic wake pattern is purely resonant and depends only on harmonics of the fundamental frequency $\Omega$.

At large separations $r\gg a$, the outgoing wave (encoded by the Hankel functions) behaves like $h_\ell^{(1)}(x)\sim (-i)^{\ell+1}e^{ix}/x\sim a/r$, which implies that the fractional density perturbation $\delta\rho(\vr,t)/\rho_0(r)$ asymptotes to a constant at large radii. Using Eq.~(\ref{eq:SIS}), we find
\begin{equation}
    \delta\rho(\vr,t) \sim \epsilon\, \rho_0(r) \quad \mbox{for}\quad r\gg a \;,   
\end{equation}
where 
\begin{equation}
    \epsilon\equiv \frac{GM}{c_s^2 a}\;.
\end{equation}      
Notice also that $M/M(<a) = \epsilon/2$, where $M(<a)$ is the mass enclosed in the profile within a radius $a$. Therefore, we must have $\epsilon\ll 1$ to ensure the validity of linear response theory, at least across a reasonable range of radii. In practice, the medium is not an ideal fluid and sound waves are damped by viscosity before they escape to infinity.

\begin{figure*}
\gridline{\fig{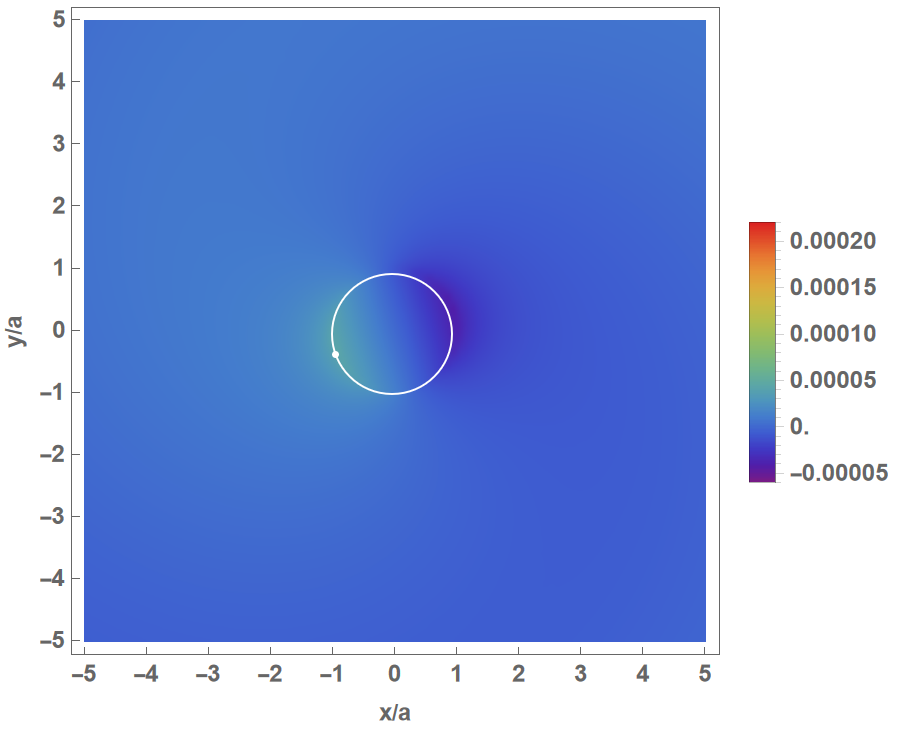}{0.45\textwidth}{}
\fig{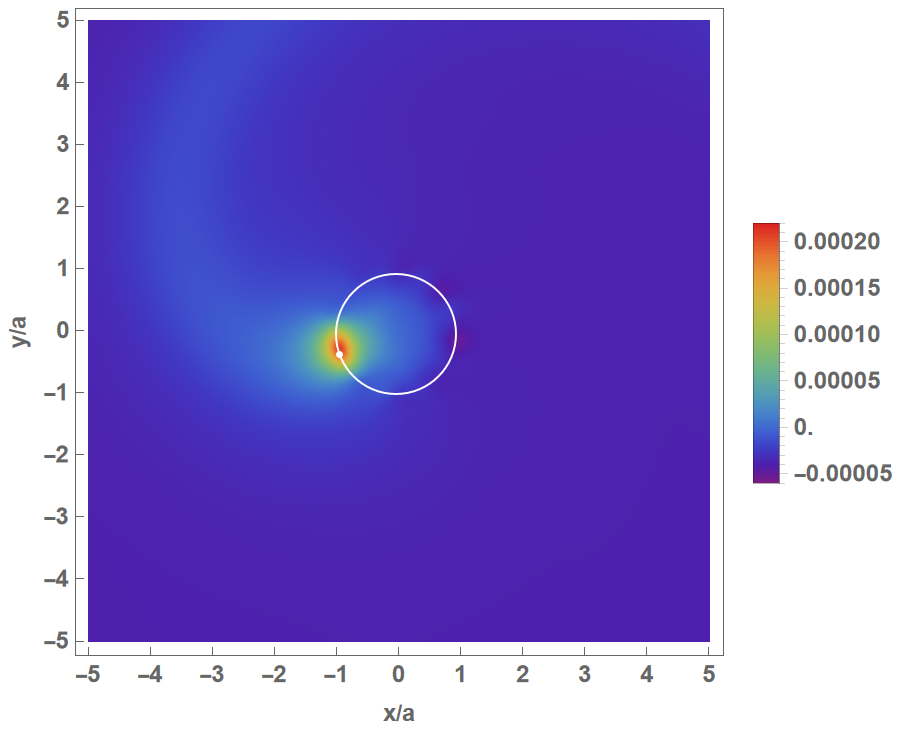}{0.45\textwidth}{}}
\gridline{\fig{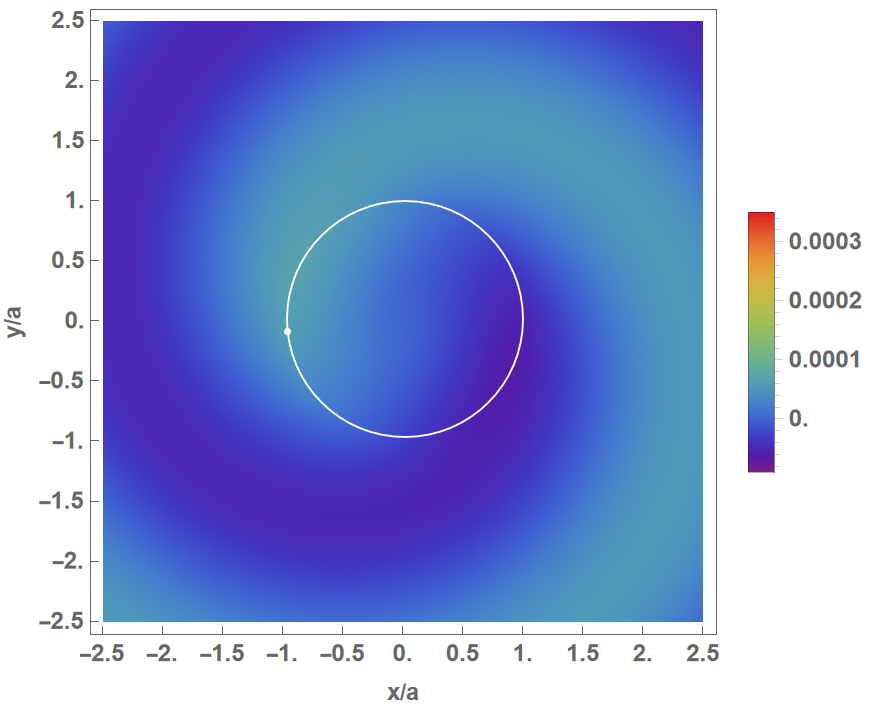}{0.45\textwidth}{}
\fig{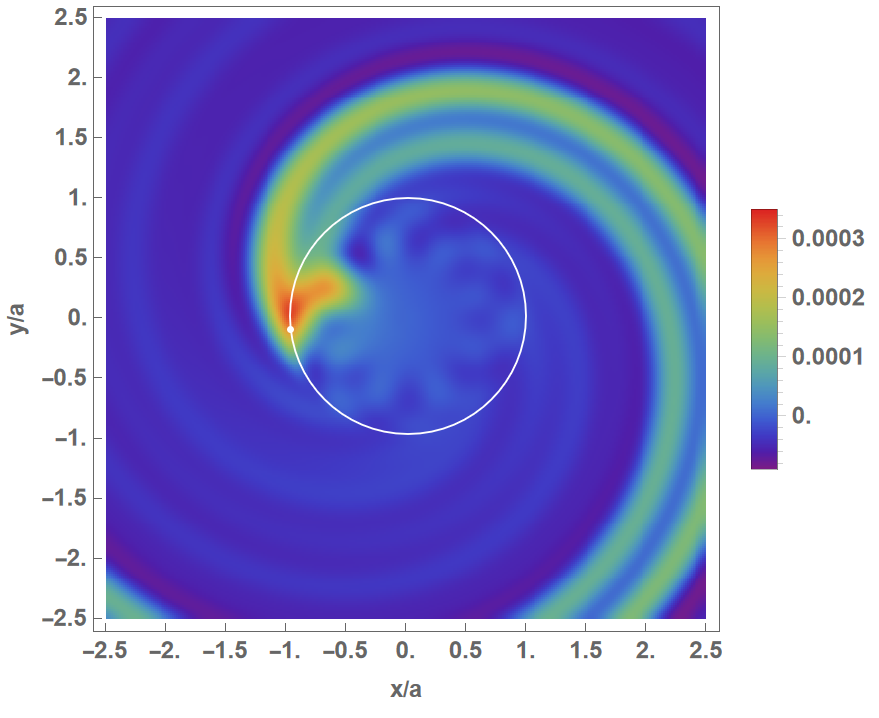}{0.45\textwidth}{}}
\caption{Steady-state acoustic wake overdensity $\delta\rho/\rho_0$ in the orbital plane of a perturber moving circularly with a Mach number of $\mach=0.2$ (left panels) and $\mach=2$ (right panels). The left panels display the dipole contribution, while the right panels show the acoustic wake for $\ell_\text{max}=8$. The mass $M$ of the perturber, the sound speed and the orbital radius $a$ are chosen such that $\epsilon=10^{-4}$ and $10^{-4}$ for $\mach=0.2$ and 2, respectively. The perturber's position is represented as a white symbol, and its circular orbit as a white circle.}
\label{fig:wake}
\end{figure*}

The monopole of the density wake follows from $A_{0,0}(r,t)$, which turns out to vanish within the orbital radius. Namely,
\begin{equation}
    A_{0,0}(r,a) = \frac{1}{2\pi G}\frac{1}{r^2}\left(\frac{1}{r}-\frac{1}{a}\right) \Theta\big(r-a\big)
\end{equation}
where $\Theta(x)$ is the Heaviside step function. As a result, the angle-average wake density profile is
\begin{equation}
    \delta\rho_{\ell=0}(r,t)=\frac{GM}{c_s^2}\rho_0(r)\left(\frac{1}{r}-\frac{1}{a}\right) \Theta\big(r-a\big)
\end{equation}
regardless of the value of $\mach$. This shows that the outer region of the profile is depleted as gas is accreted onto the perturber. By contrast, the mass $M(<a)$ enclosed in the inner region is unchanged. Note that, for $\ell\geq 1$, we have
\begin{equation}
    A_{0,\ell}(r,a) = \frac{c_s^2}{2\ell+1}\,\rho_0(r)\,\frac{r_<^\ell}{r_>^{\ell+1}} \;.
\end{equation}
Therefore, in the limit where the wake self-gravity can be neglected, the acoustic wake overdensity is
\begin{align}
    \delta\rho(\vr) &= 4\pi G M \sum_{\ell m} A_{0,\ell}(r,t)\, Y_\ell^{m*}(\rvh_p)\,Y_\ell^m(\rvh) \\
    &= \frac{GM}{c_s^2}\,\rho_0(r)\left(\frac{1}{\big|\vr-\vr_p\big|}-\frac{1}{a}\right) \quad (\mach=0) \nonumber 
\end{align}
for a static perturber. This overdensity is shown in Fig.~\ref{fig:WakeZeroMach}, together with the unperturbed and perturbed density profiles $\rho_0$ and $\rho=\rho_0+\delta\rho$, along the line of symmetry going through the center of the SIS and the perturber. A value of $GM/c_s^2a=0.1$ is assumed for illustration.

The acoustic wave patterns shown in Fig~\ref{fig:wake} illustrate the response of the self-gravitating system for supersonic and subsonic motion. Results are shown for $\mach=0.2$ (left panels) and 2 (right panels). For $\mach>1$, the acoustic wake is confined to a Mach cone, and its density diverges near the perturber's position. By contrast, a perturber moving at subsonic speed imprints a thinner, slightly overdense spiral tail analogous to the infinite, homogeneous result. The fractional density $\delta\rho/\rho_0$ converge to zero toward the center of the profile. However, one should bear in mind that $\delta\rho\sim r^{-1}$ for $r\ll a$ due to the divergence of the SIS model. The first terms in the right-hand side of Eqs.~(\ref{eq;rAwake}) and (\ref{eq:RAwake}) dominate the acoustic wake density, Eq.~(\ref{eq:drhoSIS}). They correspond to the "local source" $\rho_0\grad^{2}\delta\phi_P$ in the quantity $S(\rho_0,\phi_p,\delta\phi)$ (see Eq.~\ref{eq:soundw2}). By contrast, the remaining terms proportional to $2/(2\ell+1)$ originate from the "nonlocal source" $\grad\rho_0\cdot\grad\delta\phi_P$, and contribute to the total wake density at the $\sim 10$\% level.

\begin{figure*}
    \gridline{\fig{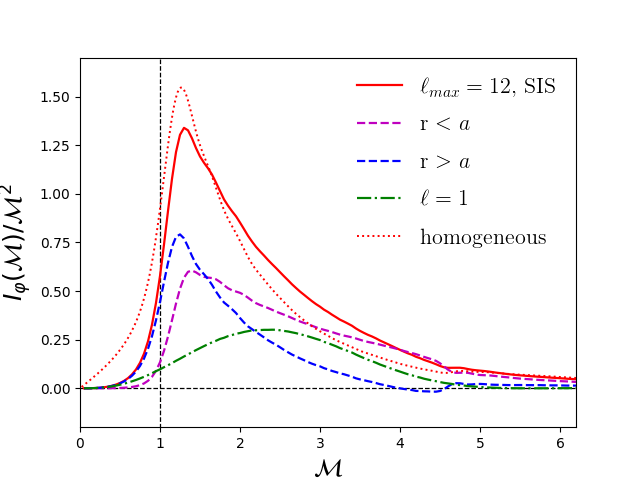}{0.5\textwidth}{}
    \fig{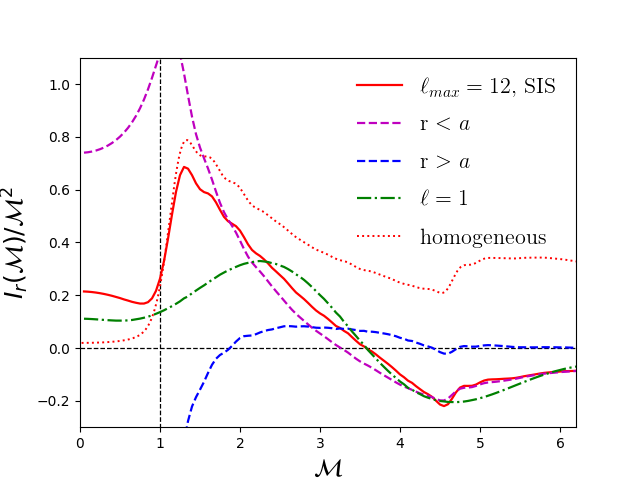}{0.5\textwidth}{}}
    \caption{Tangential (left panel) and radial (right panel) friction coefficient $I_\varphi(\mach)$ and $I_r(\mach)$ as a function of the Mach number $\mach$ in the steady-state regime for a circularly-moving perturber in a SIS profile. The dashed curves indicate the contribution from the inner and outer regions (assuming $\ell_\text{max}=12$). The (green) dot-dashed curve is the dipole contribution. For comparison, the infinite, homogeneous result is shown as the (red) dotted curve.}
    \label{fig:SIS}
\end{figure*}

\subsection{Friction coefficients}

The perturbation $\delta\phi$ to the gravitational potential of the isothermal sphere induced by a circularly-moving perturber at radius $a$ follows from Eq.~(\ref{eq:deltaphi}). First, we compute the multipole coefficients $\delta\phi_\ell^m$ assuming $r<a$. Next, we obtain the components $F_\varphi$ and $F_r$ of the DF force upon applying Eqs.~(\ref{eq:Fphi}) and (\ref{eq:Fr}), and subsequently taking the limit $r\to a$. 

The DF force can eventually be expressed as
\begin{equation}
    {\bf F}_\text{DF}(t) = -4\pi\left(\frac{GM}{\Omega a}\right)^2 \rho_0(a)\, \Big(I_r\, \rvh(t)+I_\varphi\,\pvh(t) \Big) \;.
\end{equation}
In the steady-state regime considered here, the tangential friction coefficient is given by
\begin{align}
    \label{eq:IMphiSIS}
    I_\varphi(\mach) &= \mach^2 \sum_{\ell=1}^\infty \sum_{m\ne 0} m\, \frac{(\ell-m)!}{(\ell+m)!}\,\big\lvert P_\ell^m(0)\big\lvert^2 \\
    &\qquad\times\Big[\mathcal{I}_\ell^{-}(m\mach) + \mathcal{I}_\ell^{+}\!(m\mach)\Big] \nonumber  \;.
\end{align}
The functions $\mathcal{I}_\ell^{\pm}\!(x)$ defined in Appendix \S\ref{app:technicaldetails} are related to the contribution from the inner ($-$) and outer ($+$) region. 
Likewise, the radial friction coefficient is
\begin{align}
    \label{eq:IMrSIS}
    I_r(\mach) &= i\mach^2 \sum_{\ell=1}^\infty \sum_{m} 
    \frac{(\ell-m)!}{(\ell+m)!}\, \big\lvert P_\ell^m(0)\big\lvert^2\\
    &\qquad \times \Big[(\ell+1)\,\mathcal{I}_\ell^{-}\!(m\mach) -\ell\, \mathcal{I}_\ell^{+}\!(m\mach)\Big] \nonumber \;.
\end{align}
Unlike the infinite, homogeneous case, the monopole contribution vanishes because there is no excess mass within the orbital radius. For practical evaluations, the terms in the square brackets of Eq.~(\ref{eq:IMradhom}) should be replaced by 
\begin{align}
    \label{eq:Ixto0}
    (\ell+1)\, \mathcal{I}_\ell^{-} &\to -\frac{i\,(\ell+1)}{\big(2\ell+1\big)^2} \\
    \ell\, \mathcal{I}_\ell^{+} &\to -\frac{i\,\ell}{\big(2\ell+1\big)^2}
    \nonumber 
\end{align}
when $m=0$. Appendix \S\ref{app:technicaldetails} outlines how the limiting behavior of $\mathcal{I}^\pm(x)$ can be derived. Note also that $I_\varphi(\mach)$ ($I_r(\mach)$) generally is an odd (even) function of $\mach$ since a change in the direction of the circular motion leaves $F_r$ unchanged while it flips the sign of $F_\varphi$. Therefore, a series expansion of $I_\varphi(\mach)$ ($I_r(\mach)$) around $\mach=0$ only contains odd (even) powers of $\mach$.

Fig.~\ref{fig:SIS} displays the friction coefficients $I_\varphi(\mach)$ (left panel) and $I_r(\mach)$ (right panel) as a function of the Mach number. The dashed curves indicate the contributions from the inner ($r<a$) and outer ($r>a$) region, while the dashed-dotted curve shows the contribution from the dipole. For comparison, the infinite, homogeneous results are also shown as the dotted curves. 
In the SIS case, $I_r(\mach)/\mach^2$ does not vanish in the limit $\mach\to 0$ due to the radial asymmetry of the acoustic wake around the perturber (see figure \ref{fig:WakeZeroMach}). For $\mach\ll 1$, we have
\begin{equation}
    \mathcal{I}_r(\mach) \approx \mach^2\sum_{\ell=1}^{\ell_\text{max}}\sum_m \frac{\big(\ell-m\big)!}{\big(\ell+m\big)!} \frac{\big\lvert P_\ell^m(0)\big\lvert^2}{(2\ell+1)^2}\;,
\end{equation}
which returns $I_r(\mach)\approx 0.2145 \mach^2$ for the value $\ell_\text{max}=12$ assumed in the figure. For $\mach\gg 1$, the radial friction $I_r(\mach)$ decays to zero since, unlike the infinite homogeneous case, there is no excess mass within the orbital radius. Interestingly, deviations from the infinite, homogeneous case also occur for $I_\varphi(\mach)$, which encodes the drag relevant to the dynamic of circular orbits. The striking feature is the suppression of $I_\varphi(\mach)$ in the limit $\mach\ll 1$ relative to the infinite, homogeneous expectation. Namely, the tangential friction scales as $I_\varphi(\mach)\sim \mach^5/5$, that is, $F_\varphi\propto v_p^3$ for $v_p\ll c_s$. 
This, however, would not affect the orbital decay timescale of e.g. a supermassive black hole in an $r^{-2}$ isothermal profile relative to the homogeneous case, because the circular orbit velocity $v_c=\sqrt{2}c_s$ corresponds to a Mach number $\mach=\sqrt{2}$ for which the coefficients $I_\varphi(\mach)$ are close to each other.
Finally, note that the feature at $\mach\approx 4.6033$ visible in Figs~\ref{fig:hom} and \ref{fig:SIS} corresponds to a critical Mach number at which the number of zeros of the polarization function changes \cite[see][for a related discussion]{kim/kim:2007}.

\section{Gravitational wave emission}

\label{sec:GW}

An interesting aspect of our findings is the $r^{-2}$ decay of the wake density multipoles $\delta\rho_\ell^m$ in the outer region. In a realistic astrophysical system, such a scaling would, of course, hold only over a limited range of scales. Notwithstanding, interesting effects could arise even for a truncated isothermal profile, as we will illustrate now.

\subsection{Truncated isothermal profile}

Consider a circular, binary black hole (BBH) with mass $M_2\ll M_1$, such that the massive companion $M_1$ is located at the center of the circular orbit of radius $a$ drawn by the perturber $M_2$.
This BBH is embedded in an $1/r^2$ isothermal profile extending from $\rmin$ out to a finite radial distance $\rmax$. 
The mass enclosed in a sphere of radius $a$ thus is $M(<a)=M_1+(2c_s^2/G)a=M_1(1+2/\epsilon_1(a))$. 
Here and henceforth, $\epsilon_i(a)= GM_i/c_s^2 a$ designates the ratio of (half) the Bondi radius of $M_i$ to the orbital radius. For a black hole of mass $M_1=10^6\msun$ and a sound speed $c_s=100\kms$ for example, $\epsilon_1=1$ for $a\simeq 0.43\pc$. At this orbital radius however, such a system would emit at frequencies much smaller than the nanohertz (nHz) band proved by pulsar timing arrays (PTAs). In general, $\epsilon_1\gg 1$ is required to get GW emission in the nHz frequency band, but one should relax the assumption of hydrostatic equilibrium to explore this interesting regime. Such an extension is beyond the scope of this paper, and we leave it for future work. Likewise, the validity of linear response theory requires $\epsilon_2(a)\ll 1$, which will not be satisfied for small values of $a$.

\subsection{Power radiated in GWs}

The Quadrupole formula can be applied to compute the GW emission of the BBH, had it been evolving in vacuum. In this approximation, the power radiated in GW is
\begin{equation}
P_\text{gw} = \frac{G}{5 c^5} \big\langle\dddot{Q}_{ij}\dddot{Q}_{ij}\big\rangle\;,
\end{equation}
where the temporal average $\langle \dots\rangle$ is taken over several periods, and the quadrupole moment is
\begin{equation}
    Q_{ij} = \int\!d^3r\,\delta\rho(\vr,t)\left(x_i x_j - \frac{1}{3} r^2 \delta_{ij}\right) \;.
\end{equation}
In vacuum, the circular binary would radiate GWs at a frequency $2\Omega$ with a power \citep{peters:1964}
\begin{equation}
    P_\text{gw}^\text{Vac} \approx \frac{32}{5 c^5}\frac{G^4 M_2^2 M_1^3}{a^5}
\end{equation}
for the small mass ratio assumed here. 
The effect of the truncated isothermal profile on the emission of GWs is twofold: (i) it perturbs the emission arising from the two compact bodies through a non-vacuum metric; and (ii) it provides an extra source of GWs through the time-varying quadrupole moment of the acoustic wake. Let us assess the importance of (ii) in what follows.

\begin{figure}
\includegraphics[width=9cm]{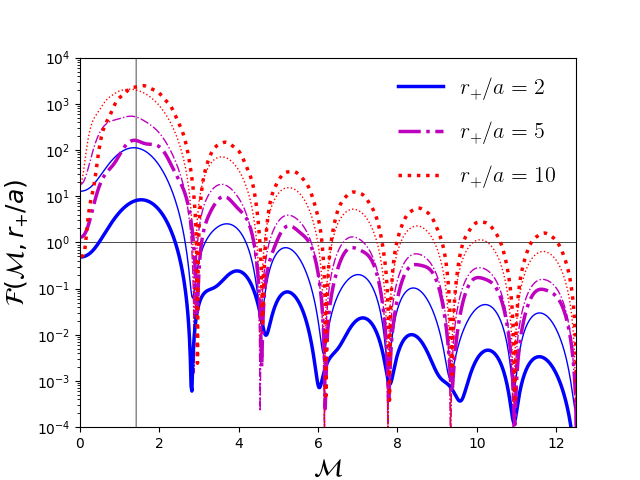} 
\caption{Shape factor $\mathcal{F}(\mach,r_+/a)$ (Eq.~(\ref{eq:FM})) as a function of Mach number for three different values of $r_+/a$ (see text for details) as quoted in the figure. The thick and thin curves represent the predictions obtained for a truncated isothermal profile and for a uniform sphere, respectively (see text for details). The vertical line indicate the Mach number $\mach=\sqrt{2}$ of a circularly-moving perturber in an isothermal profile when the latter dominates the mass of the system.}
\label{fig:FM}
\end{figure}

The acoustic wake induced by the motion of the small companion $M_2$ (the perturber) in the medium also produces GWs with a frequency $\omega_\text{gw}^2\sim GM(<a)/a^3$ or, equivalently, a (reduced) wavelength $\lambdabar_\text{gw}^2 \sim a^3 c^2/G M(<a)$. Let $r_-<a<r_+<\rmax$ be the radius of the region within which the acoustic wake has not dissipated. GW emission by the wake will be coherent only if $\lambdabar_\text{gw}/r_+\gg 1$. This condition is best satisfied if $r_+\sim \alpha a$ scales with the orbit and $\alpha \lesssim 10$. For the astrophysical system considered above for instance, taking $a=10^{-2}\pc$ gives $\lambdabar_\text{gw}\sim 4.6\pc$ is still appreciably larger than $r_+$ if $\alpha\sim$ a few. When $\lambdabar_\text{gw}/r_+\gg 1$, GW emission by the wake is coherent and, for the non-relativistic motions considered here, dominated by the quadrupole moment of the mass distribution. The latter reads
\begin{equation}
    Q_{ij} = 4\pi G M_2\sum_m \mathcal{Q}_{ij}^{(m)}\, Y_2^{m*}(\rvh_p)\, \int_{r_-}^{r_+}\!\!dr\,r^4\, A_{m\Omega,2}(r,a)
\end{equation}
where $-2\leq m\leq +2$.
The integration domain of the radial integral will be specified shortly.
Only the $m=\pm 2$ multipoles contribute to GW emission. For sake of completeness however, the matrices $\mathcal{Q}^{(m)}$ are given by $\mathcal{Q}^{(0)}=(2/3)\sqrt{\pi/5}\,{\rm diag}\big(-1,-1,2\big)$,
\begin{align}
    \mathcal{Q}^{(+1)} & =\sqrt{\frac{2\pi}{15}}\left(\begin{array}{ccc}  0 & 0 & -1 \\ 0 & 0 & i \\ -1 & i & 0 \end{array}\right) \\
    \mathcal{Q}^{(+2)} &=\sqrt{\frac{2\pi}{15}}\left(\begin{array}{ccc} 1 & -i & 0 \\ -i & -1 & 0 \\ 0 & 0 & 0 \end{array}\right) \nonumber 
\end{align}
and $\mathcal{Q}^{(-m)}=\mathcal{Q}^{(+m)*}$. The temporal average becomes 
\begin{align}
\langle \dddot{Q}_{ij}\dddot{Q}_{ij}\rangle &= \big(4\pi GM_2\big)^2 \Omega^6 \sum_{m=\pm 2}m^6 \big\lvert Y_2^m(\rvh_p)\big\lvert^2 \\
&\qquad \times \bigg\lvert\int_{r_-}^{r_+}\!\! dr\,r^4\,A_{m\Omega,2}(r,a)\bigg\lvert^2\tr\big(\mathcal{Q}^{(m)}\mathcal{Q}^{(m)*}\big) \nonumber \;,
\end{align}
which follows from
\begin{equation}
\Big\langle Y_2^{m*}(\rvh_p) Y_2^{m'}(\rvh_p)\Big\rangle = \delta_{m m'} \;.
\end{equation}
Taking advantage of the identity $\tr\big(\mathcal{Q}^{(\pm 2)}\mathcal{Q}^{(\pm 2)*}\big)=8\pi/15$, the GW power emitted by the acoustic wake in the quadrupole approximation can be expressed 
as 
\begin{equation}
P_\text{gw}^\text{Wake} = \frac{8}{15 c^5}\frac{G^4 M_2^2 M^3(<a)}{a^5}\,\mathcal{F}(\mach,r_+/a)
\end{equation} 
where 
\begin{equation}
    \label{eq:FM}
    \mathcal{F}(\mach,r_+/a) = \sum_{m=\pm 2} m^6 \frac{(2-m)!}{(2+m)!}\, \big\lvert P_2^m(0)\big\lvert^2\, \big\lvert \mathcal{A}_m \lvert^2
\end{equation}
is a "shape" factor, which depends mainly on the Mach number $\mach$ of the perturber $M_2$ and the ratio $r_+/a$ for the $r^{-2}$ profile (and the uniform medium)) considered below. There is, of course, no contribution from the time-independent term $m=0$. Explicit expressions for the (complex) quantity
\begin{equation}
    \label{eq:calA}
    \mathcal{A}_m = \frac{2\pi G}{a^2}\int^{\rmax}\!\!dr\, r^4\, A_{m\Omega,2}(r,a)
\end{equation} 
can be found in Appendix \S\ref{app:technicaldetails}.

\subsection{GW radiation from an acoustic wake}

Fig.~\ref{fig:FM} displays $\mathcal{F}(\mach,r_+/a)$ for several values of $r_+/a$ and a fixed ratio $r_-/a=0.5$. For sake of comparison, we also show the prediction for an acoustic wake propagating in a uniform sphere of radius $\rmax$. In this case, we set the density assuming $r_+=\rmax$ and equating the sound-crossing time $r_+/c_s$ to the free-fall time $(G\bar\rho)^{-1}$. This gives $\bar\rho = G^{-1}(c_s/r_+)^2$. This way, the unperturbed density is proportional to $c_s^2$ for both the truncated isothermal profile and the uniform sphere.

We have checked that the results do not change noticeably when the inner radius is set to $r_-=0$, or a region $[0.9 a, 1.1a]$ around the small companion $M_2$ is excised. In other words, all the signals originates from the outer region of the profile, which is unsurprising given the $r^{-2}$ dependence of $\delta\rho$. The vertical dotted line marks $\mach=\sqrt{2}$, which is the (constant) Mach number of a circularly-moving perturber in a truncated isothermal profile so long as $M(<a)\gg M_1$. For $\mach\approx\sqrt{2}$, the motion of $M_2$ and the acoustic wake are in resonance and maximize the shape factor $\mathcal{F}(\mach,r_+/a)$. At these Mach numbers, the ratio
\begin{equation}
     \frac{P_\text{gw}^\text{wake}}{P_\text{gw}^\text{Vac}}=\frac{1}{12}\left(\frac{M(<a)}{M_1}\right)^3\,\mathcal{F}(\mach)
\end{equation}
can be as large as $\sim 10^2$ for $M(<a)\sim M_1$ and $r_+\sim 10a$, regardless of whether the binary is embedded in an isothermal profile or a uniform medium. At fixed $a$ and $M_1$, the limit $\mach\to\infty$ is equivalent to $c_s\to\infty$, that is, a vanishing profile density $\rho_0$. In this regime, $P_\text{gw}^\text{wake}$ becomes negligible as expected. The critical density $\rho_*$ above which $P_\text{gw}^\text{wake} > P_\text{gw}^\text{Vac}$ decreases with increasing $r_+$. For $r_+\lesssim 10a$ for instance, we have $\mathcal{F}(\mach)\gtrsim 10$ for $\mach\lesssim 10$. Therefore, the condition $P_\text{gw}^\text{wake} > P_\text{gw}^\text{Vac}$ becomes $c_s^2\gtrsim 10^{-2}\Omega^2 a^2$, which implies $\rho_0(a) > \rho_*\sim 10^{-2}M_1/a^3$. This shows that, in this case, it is possible to get a significant effect even if the mass enclosed in the profile within the orbital radius is much smaller than $M_1$.

In a real fluid, acoustic wakes will be dissipated by viscosity and thermal conductivity \citep[see e.g.][]{landau/lifshitz:1975,narayan/medvedev:2001,fabian/etal:2003,sijacki/springel:2006}. Our results are valid in the limit of small viscosity, i.e. $\gamma c_s/\omega\ll 1$ where $\gamma\sim\omega^2/c_s^3\nu$ is the fluid absorption coefficient and $\omega$ is the circular frequency of the sound waves \citep{landau/lifshitz:1975}. In this case, the amplitude of the sound waves is damped according to $e^{-\gamma x}$. Since the kinematic viscosity $\nu$ approximately scales as $\nu\sim c_s \lf$, where $\lf$ is the mean free path of the fluid particles, the condition $\gamma c_s/\omega\ll 1$ is equivalent to $\lf\ll c_s/\Omega$ or, equivalently, $\lf\ll a$. Furthermore, the condition $\lf\lesssim GM_2/c_s^2$ is required for the ideal fluid approximation \citep[see e.g.][]{katz/etal:2019}.

For sound waves launched in the intracluster medium (ICM), the typical mean free path of particles due to collisions, $\lf\sim 0.1 - 10 \Kpc$ \citep{Schekochihin/cowley:2006,andradesantos/etal:2013,sanders/etal:2013}, are much larger than the orbital sizes of interest. For such systems, one should start from the Chandrasekhar formula appropriate to collisionless systems. In the (warm, diffuse) interstellar medium (ISM) of a galaxy, the typical mean free path is of order $10^{15} - 10^{16}{\ {\rm cm}}$, i.e. $\lf\sim 10^{-2} - 10^{-3}\pc$ for densities $n\sim 1\cmmm$ \citep{mckee/ostriker:2007}. For such densities however, the mass in a gaseous sphere of radius $1\pc$ is $\sim 0.1\msun$, which is too small to influence the dynamical evolution of compact binaries under consideration. This could drastically change if the compact binary is embedded in a dense gaseous cloud, a gas-rich galaxy at high redshift or, alternatively, in a dark matter core or spike. 

\section{Conclusions}

\label{sec:conclusion}

We have applied linear response theory to solve for the acoustic wake induced by a pointlike perturber in circular motion around the center of an isothermal $r^{-2}$ density profile (for simplicity). We consider the limit of an ideal fluid. Our work thus complements the analysis of \cite{tremaine/weinberg:1984,weinberg:1986}, who focused on collisionless systems. 

Our results are valid in the limit where the self-gravity of the acoustic wake can be neglected. Taking into account the wake's self-gravity leads to a (Fredholm integral) equation of the form $\delta\rho = \delta\tilde\rho + {\rm K}\,\delta\rho$, where $\delta\rho$ and $\delta\tilde\rho$  are the solution with and without self-gravity, respectively, whereas ${\rm K}$ is a nonlocal operator (due to Poisson's equation $\grad^2\delta\phi = 4\pi G\delta\rho$). Depending on the properties of ${\rm K}$, this equation could be solved iteratively (i.e. $\delta\rho=\sum_{n\geq 0} {\rm K}^n\, \delta\tilde\rho$) in analogy with the matrix method of \cite{kalnajs:1971,weinberg:1989} for collisionless systems. We have not explored this avenue here.

To calculate the acoustic wake pattern, we have expanded the polarization function in multipoles. This allows us to isolate the contribution of each multipole to the dynamical friction experienced by the perturber. The outer region $r>a$ ($a$ is the orbital radius) dominates the (tangential) drag force $F_\varphi$ for subsonic motion. For Mach numbers $\mach\ll 1$, the dipole of the acoustic wake yields the largest contribution to the drag. In this limit, the (tangential) friction coefficient scales as $I_\varphi(\mach)\sim \mach^5$ for the SIS profile, unlike the $\mach^3$ scaling in the infinite, homogeneous case. For $\mach\gg 1$ however, the two backgrounds yield similar results, in agreement with the intuition that the local approximation should always be valid deep in the supersonic regime. By contrast, predictions differ sharply for the radial component $F_r$ of dynamical friction (which is dynamically irrelevant for circular orbits). All our results assume steady-state. Comparison with numerical simulations, which requires different boundary conditions, is deferred to future work.

The circular velocity $\Omega a$ of the perturber is determined by the mass $M(<a)$ enclosed within the orbit. In the absence of a massive companion at the center, the $r^{-2}$ dependence of $\rho_0(r)$ leads to a flat rotation curve and, thereby, a constant Mach number $\mach\sim \sqrt{2}$ so long as an adiabatic sequence of quasi-circular orbit is a reasonable approximation to the inspiral of the perturber. At this Mach number, the acoustic wake pattern is in resonance with the orbital motion and maximizes the drag. Therefore, the GW emission produced by the quadrupole of the wake density pattern should also be maximized around $\mach\sim 1$. To estimate this effect, we have considered an extreme mass ratio inspiral around a massive black hole residing at the center of a (truncated) isothermal profile. Our calculation suggests that GW emission from the acoustic wake could be large relative to the vacuum predictions when $\mach\sim$ a few (i.e. in the early phase of an inspiral). These findings must be interpreted with caution as our simplified setting does not take into account neither the viscosity of real astrophysical fluids, nor the radial infall onto the massive companion. Nevertheless, it would be interesting to assess how much of this signal remains for real astrophysical systems, and the extent to which it could leave a signal in the nanohertz frequency range probed by pulsar timing arrays.

\acknowledgments

It is a pleasure to thank Shmuel Bialy, Frans van Die, Alexei Bobrick, Adi Nusser and Ivan Rappoport for discussions. G.E. would like to thank thank Anna and Alfred Gray, and Mr. Vincent Meyer for their scholarships. V.D. and R.B. acknowledge funding from the Israel Science Foundation (grant no. 2562/20).

\appendix

\section{Useful relations}

\label{app:technicaldetails}

\begin{widetext}
To derive the DF force ${\bf F}_\text{DF}(t)$ acting on a circularly-moving perturber in the steady-state regime, we calculate the perturbation $\delta\phi(\vr,t)$ to the gravitational potential assuming $r<a$ and split accordingly the radial integral Eq.~(\ref{eq:deltaphi}) into the intervals $[0,r]$, $[r,a]$ and $[a,\infty[$. The components $F_\varphi$ and $F_r$ of the DF force are obtained from Eqs.~(\ref{eq:Fphi}) and (\ref{eq:Fr}) upon taking the limit $r\to a$. In the SIS case, this last step leads to the introduction of two (complex) functions $\mathcal{I}_\ell^{\pm}(x)$ defined as
\begin{align}
    \frac{1}{x}\,{\cal I}_\ell^{-}\!(x) &= h_\ell^{(1)}\!(x)\,\mathcal{J}_\ell^{\ell+1}(x;0,1)+\frac{2}{2\ell+1}\bigg[\ell\int_0^1\!\!du\, u^{\ell+1}\,h_\ell^{(1)}\!(xu)\,\mathcal{J}_\ell^{\ell-1}(x;0,u) \label{eq:Im} \\ 
    &\qquad +\ell \int_0^1\!\!du\,u^{\ell+1}\,j_\ell(xu)\,\mathcal{H}_\ell^{\ell-1}(x;u,1) -\big(\ell+1\big)\, \mathcal{J}_\ell^{\ell+1}(x;0,1)\, \mathcal{H}_\ell^{-\ell-2}(x;1,\infty)\bigg] \nonumber \\
    \frac{1}{x}\,{\cal I}_\ell^{+}\!(x) &= j_\ell(x)\,\mathcal{H}_\ell^{-\ell}(x;1,\infty)+\frac{2}{2\ell+1}\bigg[\ell\, \mathcal{J}_\ell^{\ell-1}(x;0,1)\, \mathcal{H}_\ell^{-\ell}(x;1,\infty) \label{eq:Ip} \\ 
    &\qquad -\big(\ell+1\big)\int_1^\infty\!\!du\, u^{-\ell}\,h_\ell^{(1)}\!(xu)\,\mathcal{J}_\ell^{-\ell-2}(x;1,u) -\big(\ell+1\big) \int_1^\infty\!\!du\,u^{-\ell}\,j_\ell(xu)\,\mathcal{H}_\ell^{-\ell-2}(x;u,\infty)\bigg] \nonumber \;,
\end{align}
which encode the contribution from the inner (-) and outer (+) regions. Here, $\mathcal{J}_\ell^n(x;a,b)$ and $\mathcal{H}_\ell^n(x;a,b)$ are defined in Eq.~(\ref{eq;faux}). 

To evaluate $I_\ell^{\pm}(x)$ for $x=0$ (which arises whenever $m=0$ or $\mach=0$), we take the limit $x\to 0$ in the above expressions. This step is relatively straightforward except for the last term in the right-hand side of Eq.~(\ref{eq:Ip}). To handle it, we note that $u^{-\ell-2}h_\ell^{(1)}(u)\sim u^{-2\ell-2}$ for $u\ll 1$ and write
\begin{equation}
    x\int_1^\infty\!\!du\,u^{-\ell}\,j_\ell(xu)\,\mathcal{H}_\ell^{-\ell-2}(x;u,\infty) = x^{2\ell+1} \int_x^\infty \!\! ds\,s^{-\ell} j_\ell(s)\, s^{-2\ell-2} \bigg(s^{2\ell+2}\int_s^\infty\!\!dt\,t^{-\ell-2}\,h_\ell^{(1}\!(t)\bigg) \;.
\end{equation}
As $x$ approaches zero, the integral in the brackets converges to a constant plus subleading corrections in $s$. The constant is given by
\begin{equation}
    \lim_{s\to 0}\, s^{2\ell+1}\int_s^\infty\!\!dt\, t^{-\ell-2} h_\ell^{(1)}\!(t) = -\frac{i(2\ell-1)!!}{2(\ell+1)} \;.
\end{equation}
As a result, we have
\begin{align}
    \lim_{x\to 0} \, x\int_1^\infty\!\!du\,u^{-\ell}\,j_\ell(xu)\,\mathcal{H}_\ell^{-\ell-2}(x;u,\infty) &= \lim_{x\to 0}\, 
    x^{2\ell+1} \int_x^\infty \!\! ds\,s^{-\ell} j_\ell(s)\, s^{-2\ell-2} \bigg(-\frac{i(2\ell-1)!!}{2(\ell+1)}\bigg) \\
    &= -\frac{i}{2(\ell+1)(2\ell+1)^2} \nonumber \;.
\end{align}
The computation of the other terms in the limit $x\to 0$ proceeds analogously. Adding up all the contributions leads to Eq.~(\ref{eq:Ixto0}). 

For a truncated $r^{-2}$ isothermal profile, the (complex) function $\mathcal{A}(x)$ which appears in the shape factor $\mathcal{F}(\mach)$, Eq.~(\ref{eq:FM}), reads
\begin{align}
    \mathcal{A}(x) &= i x \left( \int_{r_-/a}^1\!\! du\, u^3\,j_2(xu)\, h_2^{(1)}(x) + \int_1^{r_+/a}\!\! du\,u^3\, j_2(x)\, h_2^{(1)}(xu) \right) \\ & \quad + \frac{2i x}{5} \int_{r_-/a}^1\!\! du\, u^3 \bigg[2 h_2^{(1)}(xu)\,\mathcal{J}_2^1(x;r_-/a,u)+2 j_2(xu)\, \mathcal{H}_2^1(x;u,1)-3 j_2(xu)\, \mathcal{H}_2^{-4}(x;1,r_+/a)\bigg] \nonumber \\
    &\quad + \frac{2ix}{5}\int_1^{r_+/a}\!\! du\,u^3 \bigg[2 h_2^{(1)}(xu)\, \mathcal{J}_2^1(x;r_-/a,1) -3 h_2^{(1)}(xu)\,\mathcal{J}_2^{-4}(x;1,u) -3 j_2(xu)\,\mathcal{H}_2^{-4}(x;u,r_+/a)\bigg] \nonumber
\end{align}
where $r_-$ and $r_+$ define the radial domain over which the acoustic wake contribute to the GW emission. For a uniform sphere of density $\bar\rho$, this becomes
\begin{equation}
    \mathcal{A}(x) = \left(\frac{2\pi G\bar\rho a^2}{c_s^2}\right) i x \bigg[h_2^{(1)}(x)\,\mathcal{J}_2^4(x;r_-,1) + j_2(x)\,\mathcal{H}_2^4(x;1,r_+/a) \bigg] \;.
\end{equation}
For the uniform density $\bar\rho = G^{-1}(c_s/r_+)^2$ assumed in \S\ref{sec:GW}, the prefactor of the last expression becomes $2\pi(a/r_+)^2$.
\end{widetext}

\bibliography{references}

\end{document}

%% file: defs.tex
\def\lsim{~\rlap{$<$}{\lower 1.0ex\hbox{$\sim$}}}
\def\bsim{~\rlap{$>$}{\lower 1.0ex\hbox{$\sim$}}}
\def\kms{\ {\rm km\,s^{-1}}}

\def\msun{\ {\rm M_\odot}}

\def\cmmm{\ {\rm cm^{-3}}}

\def\pc{\ {\rm pc}}

\def\Kpc{\ {\rm Kpc}}

\def\apjl{Astrophys.\ J.\ Lett.}

\def\aap{Astron.\ Astrophys.}

\def\apjs{Astrophys.\ J. Supp.}

\def\ln{{\rm ln}}
\def\tr{{\rm tr}}

\def\mathbi#1{\textbf{\em #1}}

\def\vtau{{\boldsymbol\tau}}

\def\pvh{\hat{\boldsymbol{\varphi}}}

\def\rvh{\mathrm{\hat{\bf{r}}}}
\def\xvh{\mathrm{\hat{\bf{x}}}}
\def\yvh{\mathrm{\hat{\bf{y}}}}
\def\zvh{\mathrm{\hat{\bf{z}}}}

\def\vr{\mathbi{r}}

\def\vv{\mathbi{v}}

\def\grad{\boldsymbol{\nabla}}

\def\mach{\mathcal{M}}
\def\rmin{r_\text{min}}
\def\rmax{r_\text{max}}

\def\lf{\lambda_\text{mfp}}